\documentclass[sn-nature]{sn-jnl}

\usepackage{graphicx}%
\usepackage{multirow}%
\usepackage{amsmath} 
\usepackage{amsthm}%
\usepackage{mathrsfs}%
\usepackage[title]{appendix}%
\usepackage{xcolor}%
\usepackage{textcomp}%
\usepackage{manyfoot}%
\usepackage{booktabs}%
\usepackage{algorithm}%
\usepackage{algorithmicx}%
\usepackage{algpseudocode}%
\usepackage{listings}%

\theoremstyle{thmstyleone}%
\theoremstyle{thmstyletwo}%

\theoremstyle{thmstylethree}%

\newcommand{\aap}{    {\it Astron. Astrophys.}}

\newcommand{\apj}{    {\it Astrophys. J.}}
\newcommand{\apjl}{   {\it Astrophys. J. Lett.}}
\newcommand{\apss}{   {\it Astrophys. Space Sci.}}

\newcommand{\jgr}{    {\it J. Geophys. Res.}}

\newcommand{\nat}{    {\it Nature}}

\newcommand{\pasa}{   {\it Pub. Astron. Soc. Aus.}}

\newcommand{\solphys}{{\it Solar Phys.}}

\newcommand{\ssr}{    {\it Space Sci. Rev.}}

\newcommand{\araa}{    {\it Ann. Rev. Astron. Astrophys.}}
\chardef\us=`\_

\begin{document}

\title[Article Title]{Observations of Successive CMEs and their Successive Type II Solar Radio Bursts in the Corona}


\author*[1]{\fnm{V.} \sur{Vasanth}}\email{vasanth.veluchamy@uj.edu.pl}



\affil*[1]{\orgdiv{}, \orgname{Astronomical Observatory of Jagiellonian University}, \orgaddress{\street{}, \city{Krakow}, \postcode{30244}, \state{}, \country{Poland}}}







\abstract{This paper reports the observations of two coronal shocks from two \textit{Coronal Mass Ejections} (CMEs) for the Successive type II Solar radio bursts observed on 02 May 2021 in the frequency range of 80 $-$ 1 MHz with the time interval of $\sim$ 20 minutes between them. Both the bursts show clear band splitting features in the harmonic band. The estimated heights for the source of the first type II burst lies in the range of 2.06 $-$ 2.93 $R_\odot$ with the average speeds of 601 $\pm$ 76, 700 $\pm$ 91 and 783 $\pm$ 105 km $s^{-1}$ for 2 X, 3.5 X and 5 X Saito electron density models, and the heights for the source of the second type II burst lies in the range of 2.24 $-$ 3.83 $R_\odot$ with the average speeds of 1063 $\pm$ 113, 1287 $\pm$ 145 and 1478 $\pm$ 172 km $s^{-1}$. The successive CMEs are observed by \textit{the twin Solar Terrestrial Relations Observatory} STEREO-A between 11:20 $-$ 12:21 UT in \textit{the Extreme Ultra Violet Imager} (EUVI) and in \textit{the Internally Occulting Refractive Coronagraphs} (COR1) FOV, the two coronal shocks are generated by the two successive CMEs observed at (11:30) 11:26 UT and (11:55:00) 11:56 UT according to ST-A (EUVI) COR1 observations and most likely released from the same active region. The average speeds of CMEs at COR1 FOV are about 574 $\pm$ 64 km $s^{-1}$ and 595 $\pm$ 82 km $s^{-1}$. The simultaneous observations of the EUV structures and the radio bursts, their coinciding height-time further confirms that the successive CMEs are responsible for the successive shocks and their related radio bursts in the corona. The observed band-splitting in the successive type-II radio bursts provides the compression ratios of 1.26 and 1.45 respectively. Therefore, these observations confirms the presence of shock waves in the corona.}




\maketitle

\section{Introduction}\label{sec1}
Type II solar radio bursts are generated by the conversion of the plasma waves excited by the electron accelerated at the MHD shocks propagating outward from the Sun, they usually appears at two frequency bands in the radio dynamic spectrum corresponding to the local plasma frequency and its harmonics \cite{bib1,bib2,bib3,bib4,bib5,bib6,bib7,bib8,bib9,bib10,bib11,bib12,bib13,bib14,bib15,bib16,bib17,bib18,bib19,bib20,bib21}. It is one of the most important diagnostic tool to understand the various eruptive phenomena in the lower solar corona. The relationship between the coronal shocks, CMEs and flares are not clear \cite{bib22}. Sometimes multiple type II bursts are observed in sequence within a time interval of $<$ 30 minutes from the first type II burst. Multiple type II bursts are first reported by Robinson and Sheridan \cite{bib23}, they concluded that a single shock wave intersecting at different coronal structures is responsible for the generation of the multiple type II bursts. While, Klassen et al \cite{bib24} supported that the second type II burst was probably driven by the evaporation shock. The characteristics and occurrence of the multiple type II bursts are studied by several authors \cite{bib25,bib26,bib27,bib28,bib29,bib30}. Most authors suggested that the multiple type II bursts were produced by a single shock either generated by the flares and/or CMEs (front or flank) travelling across the boundary of different coronal structures. While the few authors \cite{bib8,bib27,bib31} comparing the CME speed and estimated shock speed of the multiple type II radio bursts, suggested that the CME shocks at the front and flank are responsible for the first and the second type II bursts.
\hfill \break

The possibility of two different shock waves excited by one disturbance like a flare was explored by Karlicky and Odstrcil \cite{bib32}. According to the Karlicky and Odstrcil \cite{bib32}, the first shock wave is of the blast type and is of short duration compared to the second one which is piston driven by the evaporation shock. Wagner and MacQueen \cite{bib33}, suggested that the two shocks can be generated by the solar eruptions, one ahead of a CME and a blast wave moving through the CME. The possibility of a CME generating shock waves from the front and bottom was explored by Magara et al. \cite{bib34} in their model. Sakai et al. \cite{bib35} has simulated the solar type II bursts associated with the CMEs. Mancuso and Raymond \cite{bib36}, had suggested that the shock waves can be produced from the front and flanks of a CME.
\hfill \break

The present study reports the two successive type II radio bursts occurred in sequence on 02 May 2021 observed by the Nancay Decameter Array (NDA) in the frequency range of 80 $-$ 10 MHz and SWAVES in the frequency range of 16 $-$ 1 MHz, with a delay of 20 minutes between them. In Section 2, the detailed description of the observations at the EUV, white-light, X-ray and radio bands are provided, section 3 furnish the results by complying the information and features observed in the dynamics spectrum, their kinematics in the corona, as well as the kinematics of the eruptive features at the EUV and white light observations, section 4 for discussions, and section 5 conclude their results.
\hfill \break

\begin{figure}[h]
\centering
\includegraphics[width=0.95\textwidth]{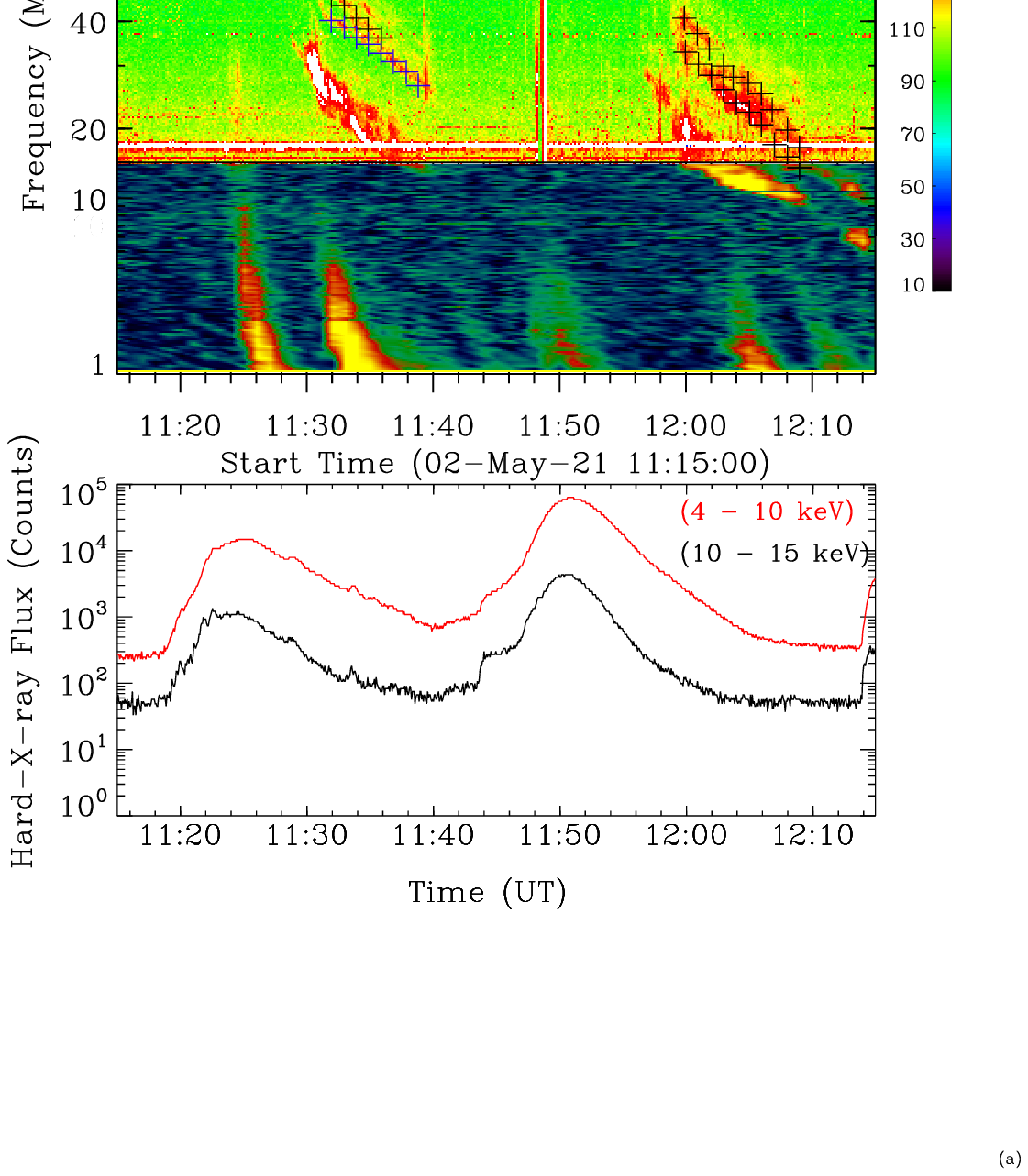}
\hfill\break
\hfill\break
\hfill\break
\caption{Radio dynamic spectrum of successive type II solar radio bursts recorded by NDA (80 $-$ 10 MHz) and SWAVES (16 $-$ 10 MHz). Both fundmental and harmonic components are clearly observed between 11:29 $-$ 11:40 UT and 11:57 $-$ 12:15 UT. The bottom panel shows the STIX hard X-ray profile in two energy bands at 4 $-$ 10 and 10 $-$ 15 keV}\label{figure1}
\end{figure}

\section{Overview of Successive Type-II solar radio bursts and their eruptive structures}\label{sec2}

The successive type-II bursts were recorded on 02 May 2021 by the Nancay Decameter Array radio spectrograph (NDA: \cite{bib37}) in the frequency range of 80 $-$ 10 MHz and the type-II burst appearing at longer wavelength continuation recorded by Wind/WAVES \cite{bib38} and STEREO-B/WAVES \cite{bib39} in the frequency range of 16 - 1 MHz (See Figure 1 for combined dynamic spectrum). The first type-II burst starts at 11:29 UT and ends at 11:40 UT, the second type-II burst starts at 11:57 UT and ends at 12:15 UT. Both the bursts presents clear band splitting structure well observed in the harmonic band.
\hfill \break

The successive type-II bursts were associated with two CMEs originating from the backside limb according to  the STEREO and the EUVI observations on board STEREO/SECCHI with one ahead and the other behind the Earth \cite{bib40}. It is important to note that STEREO-B fails by late 2014 and therefore the EUV structures and the CMEs are observed only by STEREO-A and not by STEREO-B. The longitudinal position of STEREO-A with respect to the Earth is -53 degrees, making a separation of $\sim 100$ degrees with LASCO. At larger heights the CMEs are observed by the \textit{Large Angle Spectrometric Coronagraph} (LASCO) C2 \cite{bib41} onboard the \textit{Solar and Heliospheric Observatory} SOHO Spacecraft \cite{bib42}. The observed CME structures are clear in the STEREO FOV and diffuse in other instruments. The eruptions are from the backside of the Sun and only features appears after the structures crossing the limb are observed in the \textit{Atmospheric Imaging Assembly} (AIA) onboard the \textit{Solar dynamics Observatory} (SDO) i.e., SDO/AIA and LASCO FOV. The source active region for this event (NOAA AR 12816) appears on the western side of the solar disk on 25th April 2021, several day before the event studied here. There is no GOES X-ray flare observed for the backside events. There appears double peaked hard X-ray flux profiles from the STIX observations \cite{bib43} between 11:15 $-$ 12:10 UT at 4 $-$ 10 keV and 10 $-$ 15 keV energy channels. According to the STIX hard X-ray profiles, the flux starts to rise around $\sim$ 11:19 UT, peaked around $\sim$ 11:24 UT and ends around $\sim$ 11:40 UT, similarly the second flux rise appears around $\sim$ 11:42 UT, peaked around $\sim$ 11:50 UT and ends around $\sim$ 12:10 UT. Both the type II radio burst starts at the decay phase of the hard X-ray flares (See Figure 1). The bursts starting frequencies their pattern suggested that the observed CMEs might be originated from the same active region.

\hfill \break

\section{Results}\label{sec3}

\subsection{Successive Type II Solar Radio Bursts: Spectral data at Meter wavelength}\label{subsec3}

\begin{figure}[h]
\centering
\includegraphics[width=1\textwidth]{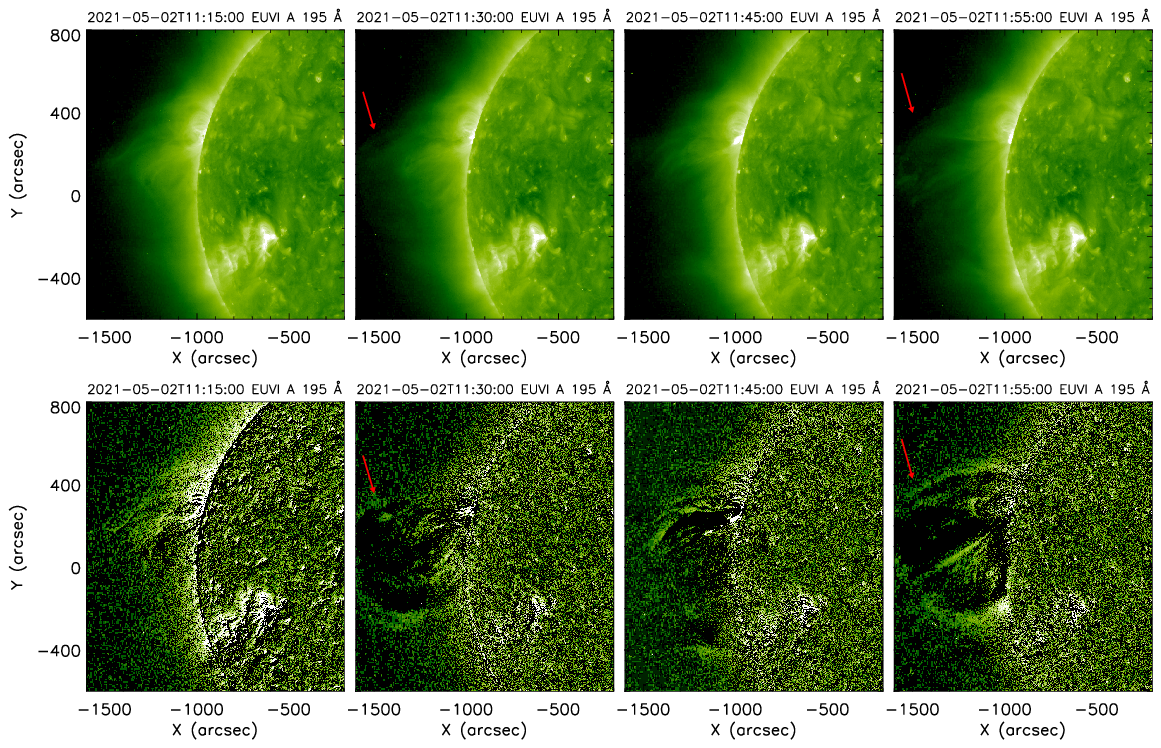}
\caption{Snapshot of ST-A EUVI images at 195 {\AA} (top panel) showing the eruptive structure and their running difference images (bottom panel). The red arrow at top and bottom panel indicates the outward propagating EUV shock waves at 11:30 and 11:55 UT. An animation of this figure is available and the duration of the animation is 4s.}\label{figure2}
\end{figure}

The combined dynamic spectrum of the successive type-II bursts observed in the frequency range of 80 $-$ 1 MHz on 02 May 2021 are shown in Figure 1. Both the type II burst shows clear fundamental and harmonic band structures as well as the band splitting features. The earlier studies in the meter wavelength radio emissions suggested that the harmonic band emissions are more intense and better defined than the fundamental bands \cite{bib44}. But for the present case, we found that the fundamental emission is more intense than the harmonic band. The first type II burst is observed in the harmonic band between 11:29 $-$ 11:40 UT, about 11 minutes in the frequency range of 47 $-$ 30 MHz. While the band-splitting is clear until 11:36 UT. Applying 2 X, 3.5 X and 5 X Saito electron density models \cite{bib45}, the heights of the type II burst are estimated by selecting several frequencies point on the harmonic band, then they are converted to the fundamental band by dividing the frequency points by 2. The estimated heights lies in the range of 2.06 $-$ 2.93 $R_\odot$, with the average speeds of 601 $\pm$ 76, 700 $\pm$ 91 and 783 $\pm$ 105 km $s^{-1}$ respectively. The first type-II burst starts at 11:29 UT and confirms the existence of shock waves, the shock like structure is observed in ST-A EUVI 195 {\AA} around 11:30 UT, and the CME is observed at ST-A COR1 around 11:26 UT. The rising loop-like feature observed by ST-A EUVI is the earliest possible CME observation detected around 11:20 UT for the first CME and around 11:42:30 UT for the second CME. The second type II burst is observed in the harmonic band between 11:57 $-$ 12:15 UT, for about 16 minutes in the frequency range of 41 $-$ 9 MHz, we used until 15 MHz were the band-splitting features are clearly identified. Again applying the 2 X, 3.5 X and 5 X Saito electron density models, the heights of type II burst are estimated and found to be in the range between 2.24 $-$ 3.83 $R_\odot$, with the average speeds of 1063 $\pm$ 113, 1287 $\pm$ 145 and 1478 $\pm$ 172 km $s^{-1}$ respectively. The second shock appears in the ST-A EUVI around 11:55 UT located at 1.68 $R_\odot$, and the CME appears at ST-A COR1 around 11:56 UT. The second CME eruption passes through the streamer region well observed in the LASCO FOV. Therefore, there is a high possibility that the high dense streamer structure acts as the radio emitting region. The successive type-II bursts were also found to be accompanied by the double peaked hard X-ray flare observed by the STIX observations.
\hfill \break

\begin{figure}[h]
\centering
\includegraphics[width=1\textwidth]{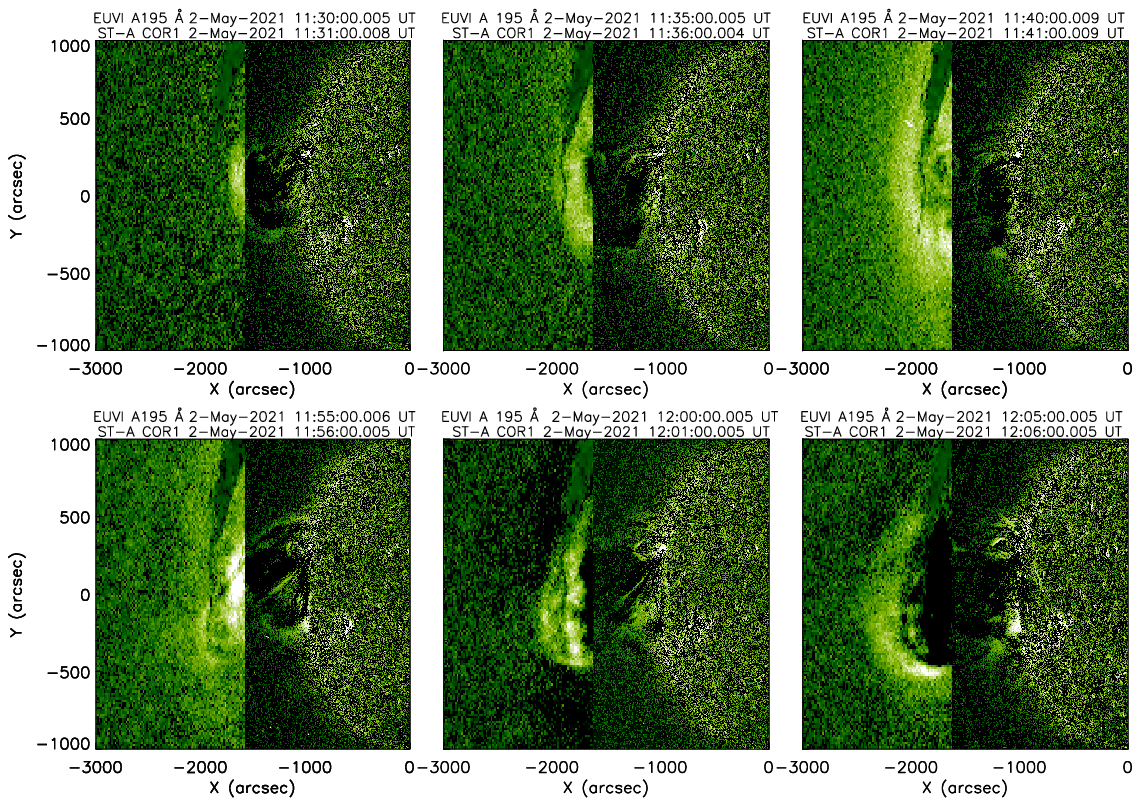}
\caption{The temporal and structural evolution of CMEs observed by ST-A COR1 FOV. Top panel shows the evolution of first CME and bottom panel for second CME.}\label{figure3}
\end{figure}

\begin{figure}[h]
\centering
\includegraphics[width=1\textwidth]{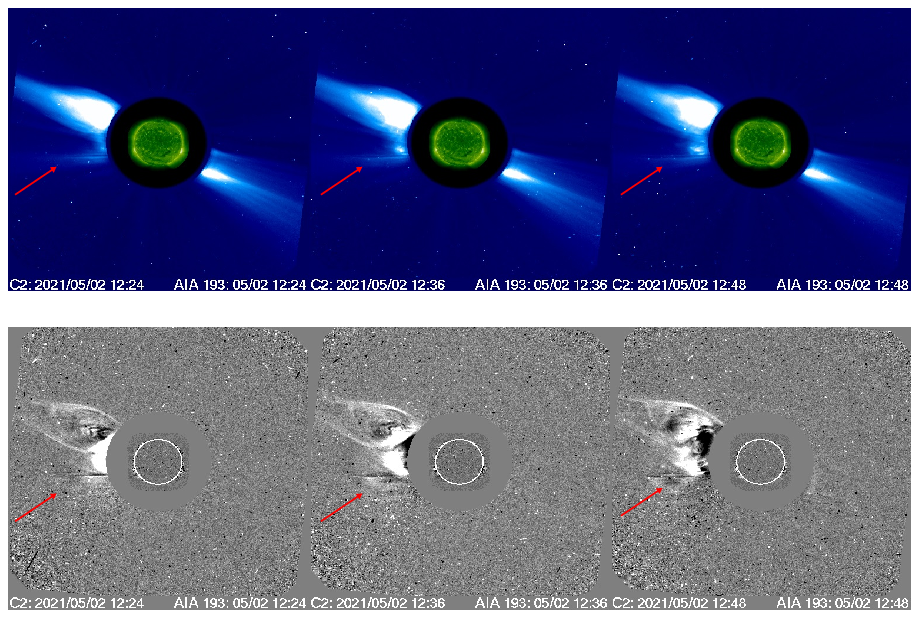}
\caption{The evolution of CMEs observed by LASCO C2 FOV. Red arrows pointing the observed CME.}\label{figure3}
\end{figure}

\subsection{EUV waves and CME structures (EUV-White Light Eruptive Structures)}\label{subsubsec3}

\begin{figure}[h]
\centering
\includegraphics[width=1\textwidth]{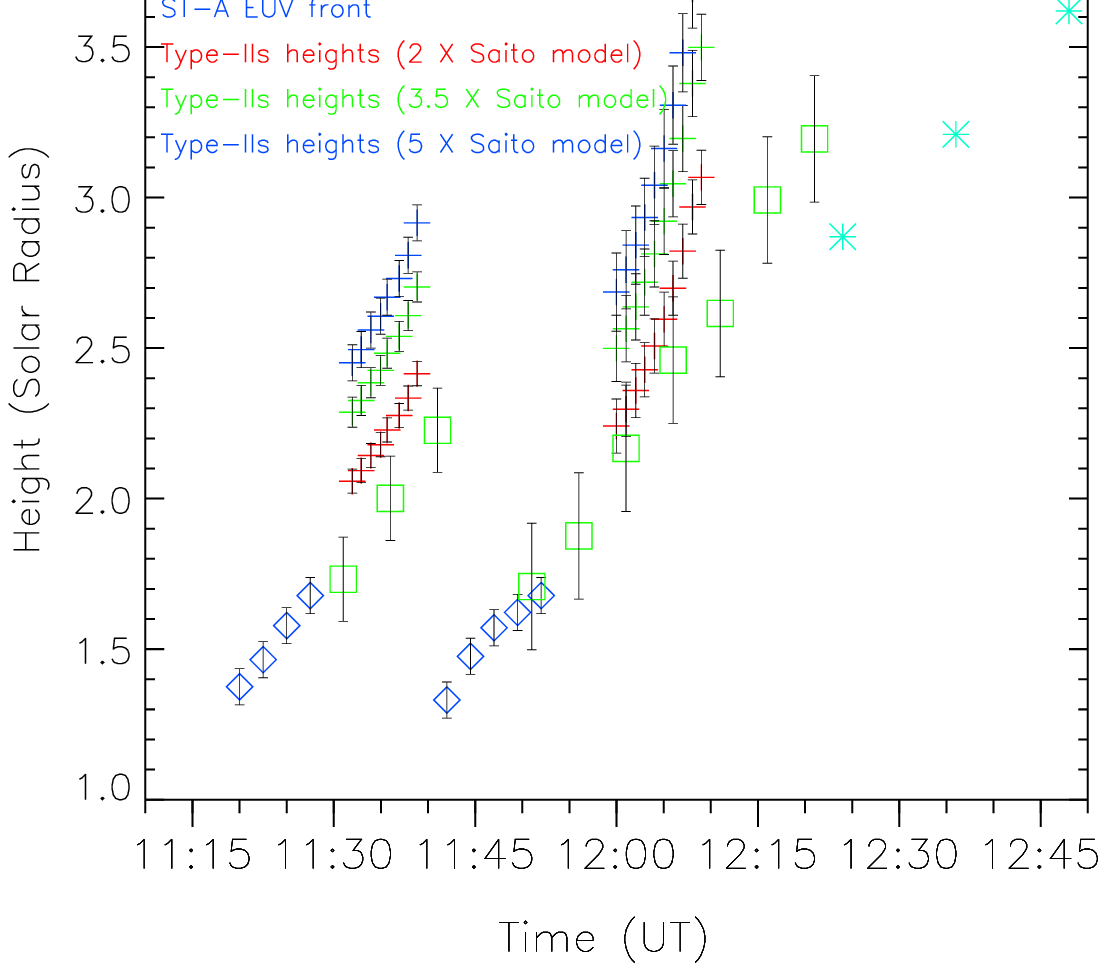}
\hfill\break
\caption{The height-time profiles observed at EUV, white--light and radio wavelengths.}\label{figure4}
\end{figure}

The relationship between EUV waves and CME structures are studied by several authors \cite{bib46,bib47,bib48,bib49,bib50}. Using the intensity enhancement profiles, Veronig et al \cite{bib47} showed the presence of a weak dome-like shock wave at ST-EUVI passbands. Ma et al \cite{bib46} reported the direct observations of the low coronal shock wave in the EUV passbands using the AIA observations. For the present study, the eruption originated on the backside of the Sun, So the AIA observations are available only after the eruption passing the limb and are also diffuse along the line of sight. The evolution of CMEs and their shock structures are observed by ST-A EUVI, ST-A COR1 and SOHO/LASCO observations (See Figure 2, Figure 3 and Figure 4). The first CME is observed in the ST-A EUVI 195 {\AA} images between 11:20 $-$ 11:35 UT, showing a slow evolution of loop system that started with the impulsive phase of the first flare around 11:20 UT. The CME structure is observed in the heights between 1.37 $-$ 1.68 $R_\odot$ in the ST-A EUVI FOV with an average speed of 468 $\pm$ 31 km $s^{-1}$, the heights are measured along the front of the CMEs and the observed type-II radio burst confirms the existence of shock wave. The CME is first observed in ST-A COR1 around $\sim$ 11:26 UT. The CME is observed in ST-A COR1 FOV between 11:26 $-$ 11:45 UT, in the height range of 1.7 $-$ 2.3 $R_\odot$, moving with an average speed of 574 $\pm$ 64 km $s^{-1}$. The evolution of the second CME is observed in the ST-A EUVI FOV between 11:42 $-$ 12:00 UT, showing a slowly evolving loop system that started with the impulsive phase of the second flare around 11:42 UT. The eruptive structure is observed in the height between 1.33 $-$ 1.68 $R_\odot$ in the ST-A EUVI FOV with an average speed of 403 $\pm$ 100 km $s^{-1}$, their earliest shock signature identified by their running difference images at 11:55 UT located at 1.68 $R_\odot$ and are observed in ST-A COR1 (11:56 UT) and LASCO C2 FOV (12:24 UT, see Figure 4) and the observed type-II radio burst confirms the existence of shock wave. The second CME is observed in ST-A COR1 FOV between 11:50 $-$ 12:30 UT, in the height range of 1.93 $-$ 3.19 $R_\odot$, moving with an average speed of 595 $\pm$ 82 km $s^{-1}$. The LASCO observation shows the presence of streamer structure and the second eruptive CME and their shock wave passes through the streamer structure and these high dense structure acts as the radio emitting region (see Figure 4). The EUV images showed the slowly propagating waves, probably the shock waves that moved outward ward and could have generated the successive type-II bursts. For the first time reporting the direct observations of two coronal shocks for the two successive type-II solar radio bursts.
\hfill \break

Figure 5 shows the height-time profiles of the radio emissions and their eruptive structures observed between 11:20 $-$ 12:45 UT across the multitude wavelength radio, EUV, and white-light observations from different instruments. The height$-$time plots from STEREO-A/SECCHI-EUVI, STEREO-A/SECCHI-COR1, and SOHO/LASCO-C2 show that the same feature is continuously tracked across the different instruments. The EUV waves from ST-A EUVI at 195 {\AA} observed between 11:20 $-$ 12:00 UT are located at 1.3 $-$ 1.68 $R_\odot$. The CMEs are observed by ST-A COR1 FOV between 11:30 $-$ 12:30 UT, in the height range of 1.73 $-$ 3.19 $R_\odot$. The radio measurements from the dynamic spectra and employing the 2 X, 3.5 X and 5 X Saito density models corresponds to the heights of 2.06 $R_\odot$ at 11:20 UT to 3.83 $R_\odot$ at 12:15 UT. Further, the height$-$time measurements yield speeds of 601$\pm$76, 700$\pm$91 and 783$\pm$105 km $s^{-1}$ for the first type II radio burst and 1063$\pm$113, 1287$\pm$145 and 1478$\pm$172 km $s^{-1}$ for the second type-II radio burst. The measured speeds of CME at ST-A COR1 FOV are $\sim$ 574 $\pm$ 64 and $\sim$ 595 $\pm$ 82 km $s^{-1}$ and in LASCO C2 FOV is 285 km $s^{-1}$. This also indicates that the successive type II radio emissions in the present case might be excited by the CME shocks.
\hfill \break

The different multipliers of Saito electron density models are used to see the variation of heights and speeds of the type-II radio bursts compared with their eruptive structures. The heights of 2 X Satio density model lies closer to the eruptive structures. The average difference in heights of 2 X Saito density model compared to 3.5 X Saito density model is 0.2 $R_\odot$ and that of the 5 X Saito density model is 0.38 $R_\odot$, and their speed differs by about 100 km $s^{-1}$ for the different multipliers of Saito electron density models.
\hfill \break



It is well known that band-splitting features in the type-II radio burst corresponds to the upstream and downstream region of the shocks \cite{bib51,bib52,bib53,bib54,bib13}. Using the relation and equations explained in \cite{bib13} and in the method section, equations 1 $-$ 7, we derived the density jump (X), Alfv{\'e}n Mach number (M$_A$), shock speed (V$_S$), Alfv{\'e}n speed (V$_A$) and magnetic field strength (B).

The evolution X and M$_A$ are presented in Figure 6. The value of X varies between 1.21 $-$ 1.37, with a mean value of 1.26 for the first type-II burst and between 1.4 $-$ 1.57, with a mean value of 1.45 for the second type-II burst. These values corresponds to the weak shock waves. The Alfv{\'e}n Mach number (M$_A$) varies between 1.16 $-$ 1.29, with a mean value of 1.2 for the first type-II burst and between 1.27 $-$ 1.45, with a mean value of 1.34 for the second type-II burst.

\begin{figure}[h]
\centering
\includegraphics[width=1\textwidth]{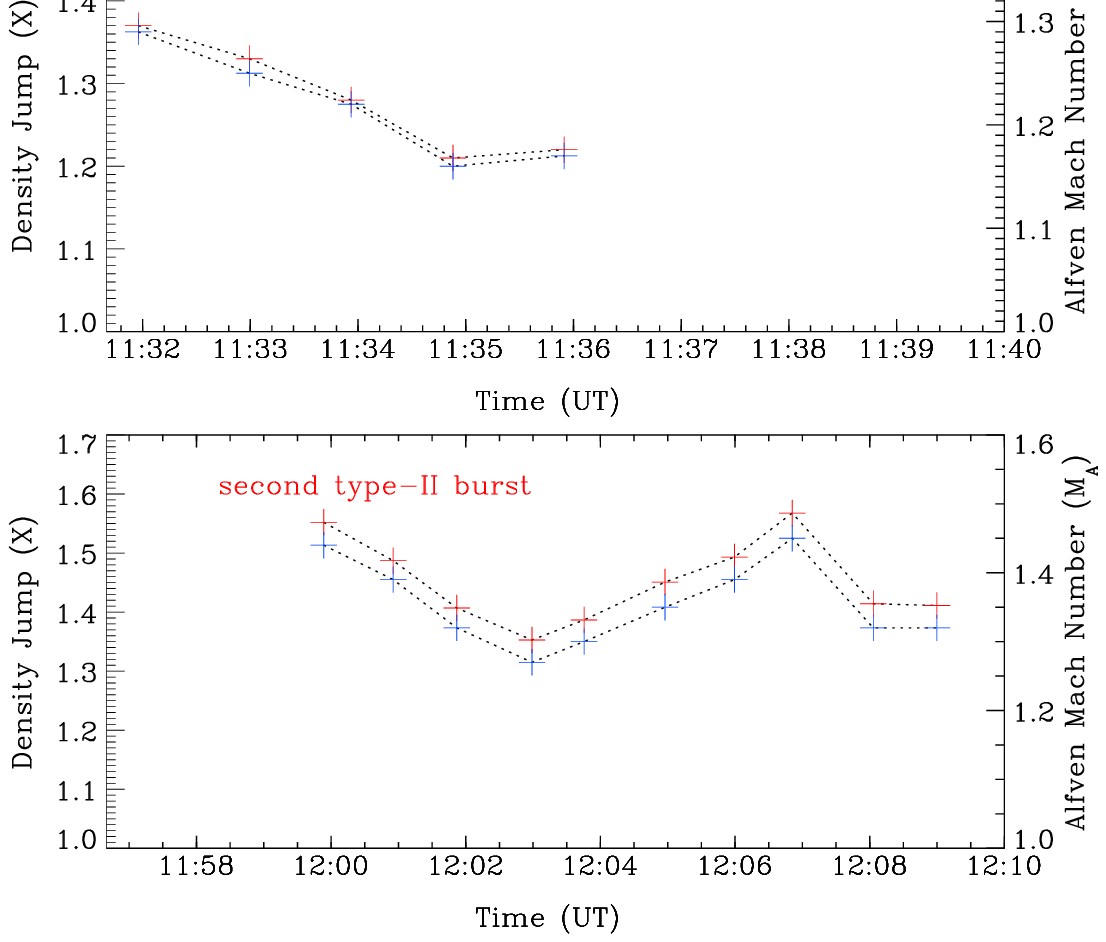}
\hfill\break
\caption{The density jump (X) and Alfv{\'e}n Mach number (M$_A$)profiles estimated from the harmonic bands of the successive type-II radio bursts.}\label{figure6}
\end{figure}

The evolution of shock speed (V$_S$)and Alfv{\'e}n speed (V$_A$) with heliocentric distance are shown in Figure 7. The coronal electron density number as a function of height ($n{_e}(R)$) decreases with height based on the employed density model. We used Saito electron density model \cite{bib45}, that is valid between 1 - 10 $R_\odot$. The mean speeds of radio source estimated by applying 2 X, 3.5 X and 5 X Saito electron density model for the first type-II burst are 601$\pm$76, 700$\pm$91 and 783$\pm$105 km $s^{-1}$ and the mean speeds of radio source for the second type-II burst are 1063$\pm$113, 1287$\pm$145 and 1478$\pm$172 km $s^{-1}$ respectively. Once we have the shock speed and Alfv{\'e}n Mach number (M$_A$), then the Alfv{\'e}n speed (V$_A$) can be estimated and the estimated mean Alfv{\'e}n speeds for the first type-II burst are 413$\pm$45,476$\pm$57 and 529$\pm$61 km $s^{-1}$ and the mean Alfv{\'e}n speeds for the second type-II burst are 789$\pm$82,955$\pm$105 and 1096$\pm$125 km $s^{-1}$.

\begin{figure}[h]
\centering
\includegraphics[width=1\textwidth]{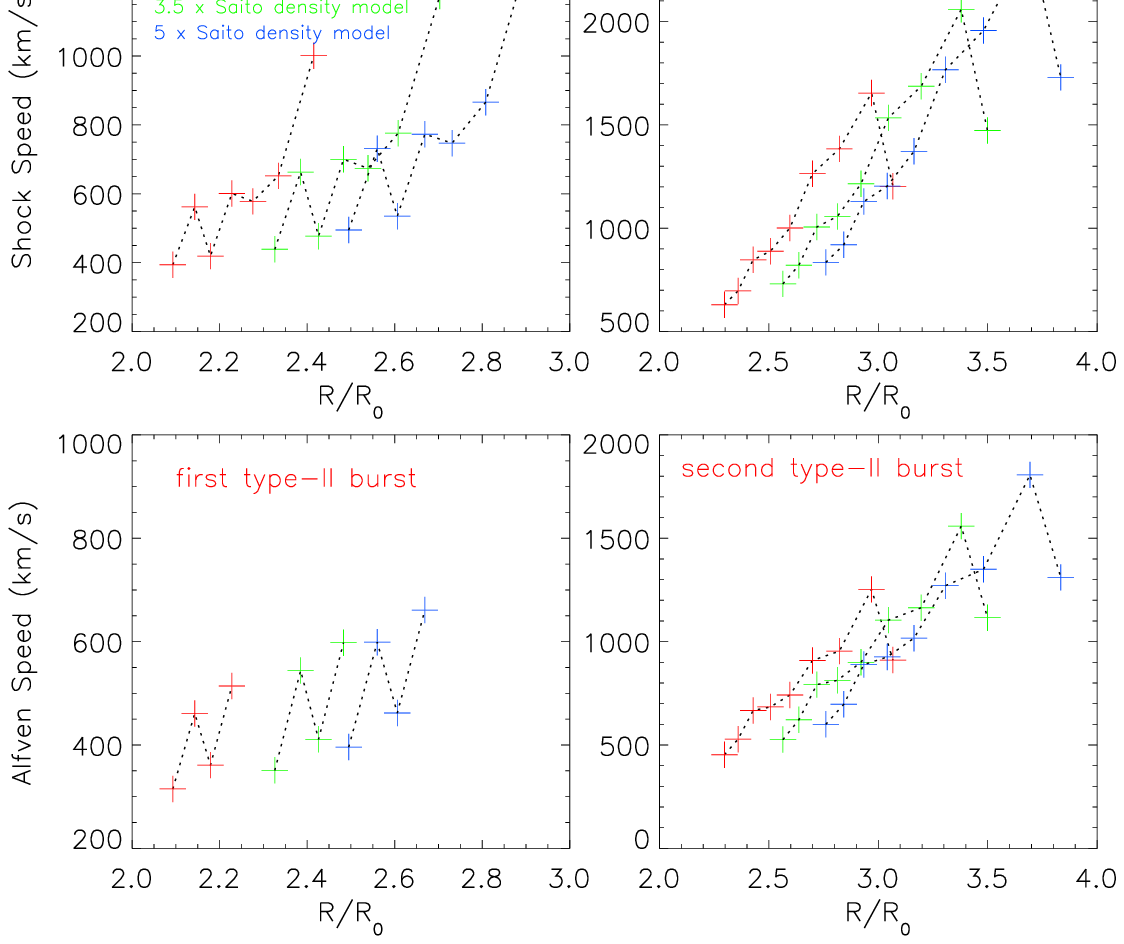}
\hfill\break
\caption{The evolution of shock speed (V$_S$) and Alfv{\'e}n speed (V$_A$) profiles estimated for the successive type-II radio bursts by employing multiplier of Saito electron density model.}\label{figure7}
\end{figure}

Figure 8 shows the magnetic-field strength profiles for the successive type-II radio bursts estimated using the Alfv{\'e}n speed and the frequency in the upstream region. The estimated magnetic field strength varies between 0.3 - 0.6 G. The mean magnetic field strengths for the first type-II burst are 0.37, 0.43 and 0.47 G and are comparable to the values obtained by Dulk and McLean \cite{bib55}. Similarly, the estimated magnetic field strength for the second radio burst varies between 0.7 - 1.5 G. The mean magnetic field strengths for the second type-II burst are 0.84, 1.02 and 1.16 G and are well above the values obtained by Dulk and McLean \cite{bib55}.

\begin{figure}[h]
\centering
\includegraphics[width=1\textwidth]{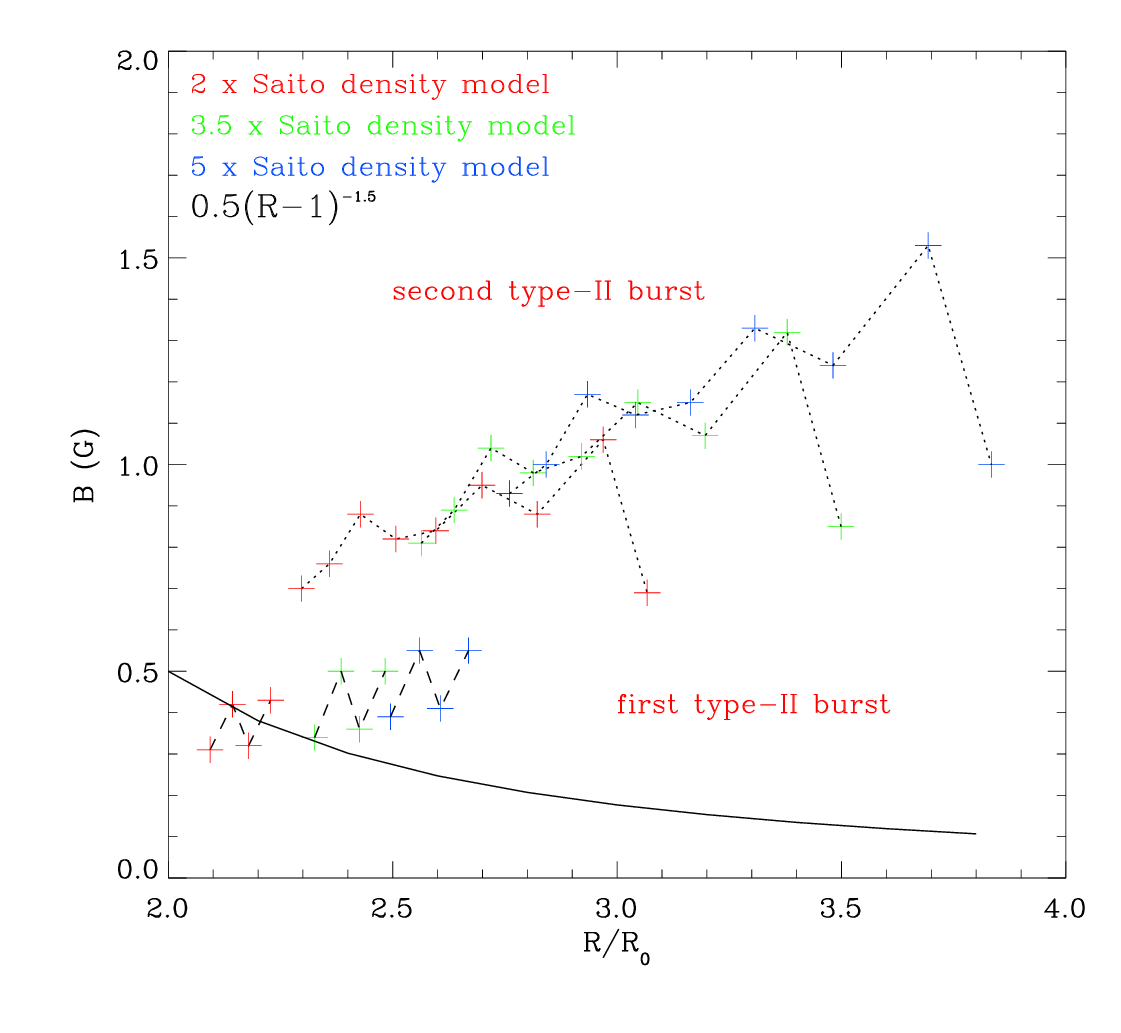}
\hfill\break
\caption{The estimated magnetic-field strength (B) profiles for the successive type-II radio burst.}\label{figure7}
\end{figure}

\section{Discussion}\label{sec4}
The present study provides the strong evidence of two coronal shocks originating from the successive CMEs generating the successive type II radio bursts. The two type II bursts were produced during the decay phase of the successive flares. Two coronal shocks were observed in ST-A EUVI (COR1) FOV around 11:30 UT (11:26 UT) and 11:55 UT (11:56 UT). The temporal and spatial evolution of the two CMEs matching with the two type II bursts, suggests the piston driven CME shocks as the possible driver for the generation of successive type II radio bursts.
\hfill \break

The multiple type II radio bursts are more than one type II burst component occurring in sequence with different frequency drift rates in the dynamic spectra. The different frequency drifts have different speeds and one of their speed will be in agreement with the CME speed \cite{bib8,bib26,bib27,bib31}. Using the theoretical studies some attempts were made to explain the generation of multiple type II bursts either by a single flare \cite{bib56} or by a CME \cite{bib34}. In such cases, they are associated with complex eruptive flare with several impulsive phases, each impulsive phase results in generating a type II radio burst, they may or may not be related to CMEs. So the multiple type II radio bursts from such cases are regarded to be generated by different energy release processes. Shanmugaraju et al. \cite{bib8} analyzed 7 multiple type II bursts that do not aligned with several impulsive phase of the flares. So they conclude that the second type II burst should be related to the CMEs.
\hfill \break

It is also important to note that some theoretical studies have suggested two different coronal shocks for the multiple type II radio bursts \cite{bib57,bib58}. But they lacks observational evidence of two different coronal shocks. Most previous observational studies support the single shock wave interacting with different coronal structures as the origin for multiple type II radio bursts. In some cases, eruptive flare with several impulsive phases during a CME may also produce multiple type II radio bursts. Therefore, most of the previous studies on multiple type-II radio bursts supports single shock waves with radio emitting energetic electron sources locating along front and flank. While for the present study, two CMEs, two shock waves are observed for the successive type-II radio bursts.


If an eruptive event has both flares and CMEs, then both flares and CMEs can produce shocks \cite{bib59, bib60,bib31,bib61,bib22}. Robinson \& Stewart \cite{bib62} suggests that flares occurring in sequence has high chance of producing multiple type II radio bursts. i.e. one near the leading edge of CME and another during the impulsive phase of the flare. Therefore the existence of two coronal shocks from two different drivers depends largely on the characteristics of the ambient medium and their conditions in the corona.
\hfill \break

\section{Conclusion}\label{sec5}

The detailed case study of the successive type-II solar radio bursts occurred on 02 May 2021 accompanied by double peaked hard X-ray flare and successive CMEs are reported. The type-IIs are observed in both the fundamental and harmonic bands. We used the harmonic band for our analysis, they are  converted to fundamental band for further height-time estimations. The estimated shock speeds by employing the 2 X, 3.5 X and 5 X Saito electron density model are found to be 601 $\pm$ 76, 700 $\pm$ 91 and 783 $\pm$ 105 km $s^{-1}$ for the first type-II burst and 1063 $\pm$ 113, 1287 $\pm$ 145 and 1478 $\pm$ 172 km $s^{-1}$ for the second type-II burst respectively. The compression ratios obtained from the band-splitting measurements for the first and second type-II bursts are 1.26 and 1.45 respectively and confirms the presence of shock waves in the corona.
\hfill \break

The two type-II radio bursts are related to two coronal shocks and their corresponding structures identified in ST-A EUVI (COR1) running difference images at 11:30 UT (11:26 UT) and 11:55 UT (11:56 UT). The estimated speed of the EUV waves are found to be 468 $\pm$ 31 and 403 $\pm$ 100 km $s^{-1}$. The evolution of CMEs are also observed in ST-A COR1 and SOHO/LASCO FOV. They appear in ST-A COR1 FOV around 11:26 $-$ 11:50 UT and later between 11:55 $-$ 12:20 UT in the height range of 1.73 $-$ 3.19 $R_\odot$. The estimated speeds of CMEs at ST-A COR1 FOV are about $\sim$ 574 $\pm$ 64 and $\sim$ 595 $\pm$ 82 Km $s^{-1}$. At larger heights, they appears in LASCO FOV after 12:24 UT, the estimated speed of CME in LASCO C2 FOV is about $\sim$ 285 km $s^{-1}$.

Comparing the height$-$time evolution of eruptive structures with the radio emissions in Figure 5, it is found that the successive type-II radio bursts in the present case is driven by the successive CMEs originated from the same active region. This is the first study to report two coronal shock structures from two successive CMEs as the origin for the successive type-II radio bursts.


\hfill \break

\section*{Data Avalability}
The datasets generated and/or analyzed during the current study are available online.\hfill \break
Nancay Decameter Array data \hfill \break
\url{https://realtime.obs-nancay.fr/dam/data_dam_affiche/data_dam_affiche.php?init=1&lang=fr}\hfill \break
STIX hard X-ray light curve data \hfill \break
\url{https://datacenter.stix.i4ds.net/view/ql/lightcurves#}\hfill \break
STEREO-A EUVI data \hfill \break
\url{https://stereo-ssc.nascom.nasa.gov/data/ins_data/secchi/L0/a/img/euvi/20210502/}\hfill \break
STEREO-A COR1 data \hfill \break
\url{https://stereo-ssc.nascom.nasa.gov/data/ins_data/secchi/L0/a/seq/cor1/20210502/}\hfill \break

\section*{Acknowledgements}

This work is supported by the POB Anthropocene research program of Jagiellonian University, Krakow, Poland. I would like to thank Dr.V.V.Grechnev and Dr.Bing Wang for the fruitful discussion and help in making images. The author greatly acknowledges various online data centers of NOAA and NASA for providing the data. We express our thanks to the Nancay Decameter array (NDA), STIX, STEREO/SECCHI, and SOHO/LASCO teams for providing the data and LASCO CME catalog generated and maintained by the Center for Solar Physics and Space Weather, the Catholic University of America, in cooperation with the Naval Research Laboratory and NASA.
\hfill \break

\section*{Author contribution}

V.V selected the research problem, performed the data analysis, interpretation of results and prepared the manuscript.

\section*{Conflict of interest/Competing interests}
The author declare no competing interests.


\section*{Methods}
\subsection*{ Type II Solar Radio Bursts speed estimation}

The frequency of plasma emission, $f_p$, is directly proportional to the electron plasma density and can be expressed by the following relation:
\begin{equation}
f_p = 9000 \sqrt n_e MHz
\end{equation}

For radio emission at the plasma frequency (f = $f_p$), the frequency drift rate df/dt, can be used to determine the speed (V), of the electron beam generating plasma emission using the following relation:
\begin{equation}
V = \frac{2\sqrt n_e}{C} (\frac{dn_e}{dr})^{-1} \frac{df}{dt}
\end{equation}

The unknown parameter in the above equation is plasma density ($n_e$).The electron density abruptly decreases in the corona with increase in height, therefore it is necessary to consider an electron density model, $n_e$ = $n_e$(r), to obtain the density at a specific height. The Saito electron density model \cite{bib43} is used to obtain the speeds of multiple type II radio bursts in the 'Successive Type II Solar Radio Bursts: Spectral data at Meter wavelength' section. The drift rate (df/dt) of the radio bursts were estimated directly from dynamic spectra by computing the ratio between frequency and total duration of each type II radio burst.
\break

Therefore, the type II bursts observed at a local plasma frequency (f) or its harmonics can be used to get information about the plasma density (n).
The region ahead of the shock (upstream region) is characterized by the electron density (n1) and plasma frequency (f1). The emission from this region creates the lower-frequency branch (LFB) of the split band. The region behind the shock (downstream region) is compressed, so the electron density n2 is higher than n1 and the corresponding plasma frequency f2 is higher than f1. The emission from the downstream region corresponds to the upper frequency branch (UFB) of the split band.

The relative instantaneous bandwidth (BDW) of the splitting can be expressed as
\begin{equation}
BDW = \Delta f/f = (f2 - f1)/f1 = (n2/n1)^{1/2} -1
\end{equation}
Thus, the density jump (X) across the shock can be written as
\begin{equation}
X= n2/n1 = (BDW+ 1)^2
\end{equation}
From the density jump (X) we derive the Alfv{\'e}n Mach number M$_A$ using a simplified Rankine–Hugoniot jump relation,
\begin{equation}
 M_A = (X(X +5)/2(4 -X)))^{1/2}
\end{equation}
which is valid for perpendicular shocks in a low plasma-to-magnetic pressure ratio environment ($\beta \ll$ 1) and $\gamma$ = 5/3.
We used the emission frequency and the drift rate of the harmonic band to estimate the
height of the radio source and the shock speed, employing the appropriate coronal electron-density
models. Once the shock speed (V$_{s}$) is estimated, it is possible to convert the Alfv{\'e}n Mach
number to the Alfv{\'e}n speed using the relation
\begin{equation}
V_{A} = V_{s} / M_{A}
\end{equation}
We determine the ambient magnetic field strength (B) using the Alfv{\'e}n speed (V$_A$) and the LFB frequency (f1):

\begin{equation}
B = 5.1*10^{-5} * f_1 V_{A}
\end{equation}

where the frequency is expressed in megahertz, the Alfv{\'e}n velocity in kilometers per second, and the magnetic field in gauss.

\end{document}